\documentstyle[preprint,aps,floats,epsfig]{revtex}
\tighten

\def\beq{\begin{equation}}
\def\eeq{\end{equation}}
\def\beqn{\begin{eqnarray}}
\def\eeqn{\end{eqnarray}}
\def\nn{\nonumber\\}

\def\vec#1{\mbox{\boldmath $#1$}}  
\def\GeV{{~\rm GeV}}

\begin{document}

\preprint{submitted to Phys. Lett. B}

\title{Sum rule for the backward spin polarizability of the nucleon \\
from a backward dispersion relation}

\author{A.I. L'vov$^1$ and A.M. Nathan$^2$}

\address{
\medskip
$^1$P.N. Lebedev Physical Institute of the Russian Academy of Sciences,
  \\ Leninsky Prospect 53, Moscow 117924, Russia \\
\medskip
$^2$Department of Physics, University of Illinois at Urbana-Champaign,
  \\ Loomis Laboratory of Physics, 1110 West Green Street, Urbana,
   IL 61801-3080, USA}

\maketitle

\begin{abstract}

A new sum rule for $\gamma_\pi$, the backward spin polarizability of
the nucleon, is derived from a backward-angle dispersion relation.
Taking into account single- and multi-pion photoproduction in the
$s$-channel up to the energy $\omega_{\rm max} = 1.5\GeV$ and
resonances in the $t$-channel with mass below $1.5\GeV$, it is found
for the proton and neutron that $[\gamma_\pi]_p = -39.5 \pm 2.4$ and
$[\gamma_\pi]_n = 52.5\pm 2.4$, respectively, in units of $10^{-4} ~\rm
fm^4$.

\end{abstract}

\section{Introduction}

With recent progress in developing and using effective field theories,
a practical knowledge of various low-energy parameters of hadrons and
their interactions is of high current interest.  Among such parameters
are $\gamma_i$, the four spin polarizabilities of the nucleon, which
characterize the spin-dependent response of the internal degrees of
freedom to external soft electromagnetic fields
\cite{ragu93,bern95,hols98,drec98,babu98}.  Two linear combinations of
$\gamma_i$ have an especially transparent meaning.  These are the
forward and backward spin polarizabilities, $\gamma$ and $\gamma_\pi$,
which are defined as the coefficients of $i\omega^3 \vec\sigma \cdot
\vec e^{\prime *} \times \vec e$ in the structure-dependent (non-Born)
part of the low-energy forward or backward Compton scattering
amplitude.  From the Gell-Mann--Goldberger--Thirring dispersion
relation, one can predict the forward spin polarizability $\gamma$
through the total photoabsorption cross sections with polarized beam
and target with total helicities 1/2 and 3/2 \cite{bern92}:
\beq
\label{gamma}
  \gamma = \int_{\omega_0}^\infty
   (\sigma_{1/2} - \sigma_{3/2})\,\frac{d\omega}{4\pi^2\omega^3}.
\eeq
Using more complicated dispersion relations, two other $\gamma$'s can
also be found \cite{drec98,babu98}.  However, this approach is not
sufficient to predict the backward spin polarizability which is
particularly sensitive to that part of the high-energy behavior of the
spin-dependent Compton scattering amplitude driven by the invariant
amplitude $A_2$ \cite{babu98,lvov97}. A dispersion estimate for
$\gamma_\pi$ obtained in Refs.~\cite{drec98,babu98} using an
unsubtracted fixed-$t$ dispersion relation for $A_2$ was based on the
strong assumption that the high-energy asymptotics of $A_2$ are
entirely determined by $\pi^0$ exchange \cite{lvov97}, whereas other
possible exchanges with heavier mesons or few-pion states are
negligible.  Although not well justified, this estimate still gives a
result close to that obtained in the framework of leading order
chiral perturbation
theory (ChPT), with the $\Delta$-isobar included
through the small-energy-scale expansion \cite{hols98}.  In particular,
it was found for the proton
\beq
\label{th}
   [\gamma_\pi]_p =
     -36.7~\cite{hols98}, \quad
     -34.3~\cite{drec98}, \quad
     -37.2~\cite{babu98}
\eeq
(hereafter the units used for $\gamma$'s are $10^{-4} ~\rm fm^4$).  In
all these calculations, the magnitude of $\gamma_\pi$ is dominated by
the contribution to nucleon Compton scattering from the $t$-channel
$\pi^0$-exchange which yields
\beq
\label{pi0}
  \gamma_\pi^{(\pi^0)} =
   \frac{g_{\pi NN}(0) F_{\pi\gamma\gamma}(0)} {2\pi m_{\pi^0}^2 m}
   \tau_3 = (-45.0\pm 1.6)\tau_3,
\eeq
where $m$ is the nucleon mass and $\tau_3$ is equal to 1 or $-1$ for
the proton or neutron, respectively.  Experimental $\pi^0\gamma\gamma$
and $\pi NN$ couplings extrapolated to $t=0$ were taken to obtain the
numerical value of $\gamma_\pi^{(\pi^0)}$.  In the framework of ChPT,
$\gamma_\pi^{(\pi^0)} = -e^2 g_A \tau_3/ (8\pi^3 f_\pi^2 m_{\pi^0}^2) =
-45.3 \tau_3$.

Recently, these theoretical predictions have been challenged by the
first experimental estimate of the backward spin polarizability of the
proton \cite{legs98} which was (indirectly) obtained from a
simultaneous analysis of pion photoproduction and unpolarized Compton
scattering data:
\beq
\label{exp}
   [\gamma_\pi]_p = -27.1 \pm 3.4,
\eeq
where statistical, systematic, and model-dependent errors were added in
quadrature. Isolating the well-defined contribution (\ref{pi0}) with
errors already included in Eq.~(\ref{exp}), one can infer the
non-$\pi^0$ part of the backward spin polarizability which is of the
most theoretical interest:
\beq
\label{exp0}
  [\gamma_\pi^{(\rm non-\pi^0)}]_p = 17.9 \pm 3.4.
\eeq
On the other hand, all the cited theoretical approaches yield roughly
one half of Eq.~(\ref{exp0}), with $\simeq +4$ coming from nonresonant
pion production and another $\simeq +4$ from the $\Delta$-resonance
excitation. Therefore, the experimental finding (\ref{exp0}) suggests
that there is another missing and very large positive contribution to
$\gamma_\pi$.  In the framework of ChPT, this missing contribution
might be due to next-to-leading-order effects, especially in the
related counter-terms.  In the framework of fixed-$t$ dispersion
relations, the missing contribution could be sought in the high-energy
part of the poorly-convergent dispersion relation for the amplitude
$A_2$ or in heavy-meson exchanges which might contribute to $A_2(s,t)$
at large $s$ and $t=0$.

Neither an extension of the ChPT calculations to higher orders nor a
more exact treatment of the high-energy behavior of $A_2$ offer much
promise to resolve this issue, since both approaches are technically
difficult and suffer from badly-controlled uncertainties.  Therefore we
develop here a different approach based on a backward-angle dispersion
relation which is manifestly free from the convergence problem. This
approach is very similar to that proposed for determining the
difference of dipole electric and magnetic polarizabilities of the
nucleon \cite{bern74}.  We derive a sum rule for $\gamma_\pi$ with
well-defined $s$- and $t$-channel contributions and then use it to
predict the backward spin polarizability of the nucleon.

\section{Sum rule}

We start by recalling the form of dispersion relations for nucleon
Compton scattering amplitudes at the fixed scattering angle
$\theta=\pi$ \cite{bern74,hear62}. At this specific angle, the
corresponding Mandelstam variables obey two constraints, $s+u+t=2m^2$
and $su=m^4$, and the structure of physical cuts in the complex plane
of $s$ is particularly simple.  When $\theta=\pi$, we have for any
cross-even analytical function $A(s,u,t)=A(u,s,t)$ which vanishes at
high $s$ and has singularities only at physical thresholds in $s$, $u$,
or $t$ channels:
\beqn
\label{DR}
  {\rm Re}\,A(s,u,t) &=& A^{\rm pole}(s,u,t) +
     \frac1\pi {\rm P}\int_{s_0}^\infty
   \Big( \frac{1}{s'-s} + \frac{1}{s'-u} - \frac{1}{s'} \Big) \,
   {\rm Im}_s A(s',u',t')\,ds'
\nn && \qquad {}
    + \frac1\pi {\rm P}\int_{t_0}^\infty
     {\rm Im}_t A(s',u',t')\frac{dt'}{t'-t},
\eeqn
where $s'+u'+t'=2m^2$ and $s'u'=m^4$.  ${\rm Im}_s$ and ${\rm Im}_t$
denote the imaginary parts of the amplitude in the $s$ and $t$ channels,
which start at the thresholds $s_0=(m+m_\pi)^2$ and $t_0=m_\pi^2$,
respectively.

We apply the backward dispersion relation (\ref{DR}) to the function
\beq
\label{A}
   A \equiv A_2 + \Big(1-\frac{t}{4m^2}\Big)A_5,
\eeq
where $A_i(s,u,t)$ are cross-even invariant amplitudes of Compton
scattering \cite{babu98,lvov97} free from kinematical singularities and
constraints.  The function $A$ determines the spin-dependent part of
the Compton scattering amplitude $T_{fi}$ in the backward kinematics,
which has the following form in the Lab frame \cite{babu98}:
\beqn
  \frac{1}{8\pi m} \Big[T_{fi}\Big]_{\theta = \pi}
  &=&  -\frac{\omega\omega'}{2\pi} \sqrt{1-\frac{t}{4m^2}}
   \; \vec e^{\prime *}\cdot \vec e\,
       \Big( A_1-\frac{t}{4m^2}A_5 \Big)
\nn && \qquad
   - \frac{\omega\omega'}{2\pi m} \sqrt{\omega\omega'}
     \; i\vec\sigma \cdot \vec e^{\prime *} \times \vec e\,
    \Big[ A_2+\Big(1-\frac{t}{4m^2}\Big)A_5 \Big].
\eeqn
Here $\omega$ and $\omega'=\omega (1+2\omega/m)^{-1}$ are energies of
the initial and final photons, so that $s=m^2+2m\omega$,
$u=m^2-2m\omega'$, and $t=-4\omega\omega'$.

Assuming for the amplitude $T_{fi}$ a standard Regge behavior $\sim
s^{\alpha_R(u)}$ for high $s$ and fixed $u$, we have
\beq
   A(s,u,t) \sim s^{\alpha_R(0)-3/2} \quad
   \mbox{when~} s\to\infty, ~u\to 0.
\eeq
Here $\alpha_R(0)=\frac32 - \alpha' m_\Delta^2 \simeq 0.13$ for the
leading Regge trajectory, which is that of the $\Delta(1232)$ resonance
with the slope $\alpha'\simeq 0.9\GeV^{-2}$.  Such a high-energy
behavior of $A$ guarantees convergence of the dispersion relation
(\ref{DR}).

In the Born approximation, which is determined by the electric charge
$q=(\tau_3+1)/2$ and the anomalous magnetic moment $\kappa$ of the
nucleon, the amplitude $A$ becomes a (double) pole function of $s$,
\beq
   A^{\rm Born}(s) =
   me^2\frac{\kappa^2 + 4q\kappa + 2q^2}{(s-m^2)(u-m^2)}
\eeq
(here $e^2\simeq 4\pi/137$ and $su=m^4$). It vanishes at high $s$ and
hence coincides with $A^{\rm pole}$ in the dispersion relation
(\ref{DR}).  Correspondingly, the integrals in Eq.~(\ref{DR}) give the
non-Born part $A^{\rm NB}= A - A^{\rm Born}$ of the function $A$.  When
$s=u=m^2$ and $t=0$, these integrals determine the backward spin
polarizability of the nucleon,
\beq
   \gamma_\pi \equiv -\frac{A^{\rm NB}(m^2,m^2,0)}{2\pi m}
        =  \gamma_\pi^s  + \gamma_\pi^t,
\eeq
where
\beqn
\label{s}
    \gamma_\pi^s &=&   -\frac{1}{2\pi^2 m} \int_{s_0}^\infty
      \frac{s'+m^2}{s'-m^2} \,  {\rm Im}_s A(s',u',t')\,\frac{ds'}{s'}, \\
\label{t}
    \gamma_\pi^t &=&   -\frac{1}{2\pi^2 m}\int_{t_0}^\infty
     {\rm Im}_t A(s',u',t')\frac{dt'}{t'}.
\eeqn
We will refer to Eqs.~(\ref{s}) and (\ref{t}) as $s$- and $t$-channel
contributions, though actually $\gamma_\pi^s$ includes contributions
from both $s$ and $u$ channels.  Applying unitarity and using the
well-known formalism of helicity amplitudes, one can express ${\rm
Im}_s A$ in Eq.~(\ref{s}) in terms of the photoabsorption cross
sections $\sigma_\lambda^n$ with specific total helicity $\lambda$ of
the beam and target and with relative parity $P_n=\pm 1$ of the final
state $n$ with respect to the target, resulting in
\beq
\label{SR}
   \gamma_\pi = \int_{\omega_0}^\infty   \sqrt{1+\frac{2\omega}{m}}
    \Big(1+\frac{\omega}{m}\Big)  \sum_n P_n \Big(
    \sigma_{3/2}^n(\omega) - \sigma_{1/2}^n(\omega) \Big)
    \frac{d\omega}{4\pi^2\omega^3} + \gamma_\pi^t.
\eeq
This sum rule for the backward spin polarizability of hadronic spin-1/2
systems is our main result.

The $s$-channel contribution to $\gamma_\pi$ is given by a manifestly
convergent integral of spin-dependent partial cross sections which are
bound by the total cross section, $\sum_n (\sigma_{3/2}^n +
\sigma_{1/2}^n) = 2\sigma_{\rm tot}$.  The integrand in Eq.~(\ref{SR})
has a parity structure similar to that found for the spin-independent
part of $T_{fi}$ \cite{bern74}:
\beq
\label{ds}
  \Delta\sigma \equiv \sum_n P_n (\sigma_{3/2}^n - \sigma_{1/2}^n) =
  \Big\{ \sigma_{1/2}(\Delta P={\rm yes}) -
   \sigma_{1/2}(\Delta P={\rm no}) \Big\} -
  \Big\{ 1/2 \to 3/2 \Big\}.
\eeq
In comparison with Eq.~(\ref{gamma}), photoexcitations $n$ which carry
the same parity as the nucleon (i.e., $\Delta P={\rm no}$) contribute
to the backward spin polarizability with an opposite sign.

The $t$-channel contribution (\ref{t}) cannot be recast through
directly observable quantities, because a large part of the integration
region is unphysical, $t<4m^2$. Nevertheless the following general
remarks provide guidance for the nature of this contribution.  The key
point is that the imaginary part in Eq.~(\ref{t}) is determined by
intermediate states of even total angular momentum $J$, charge parity
$C=+1$, isospin $I\le 2$ ($I\le 1$ for the nucleon target), and
ordinary parity $P=-1$; i.e.,\
\beq
\label{jpc}
   J^{PC}=0^{-+}, ~2^{-+}, ~4^{-+}, ~\ldots ~, \quad I=0,1,2.
\eeq
The property $P=-1$ can be easily seen from the relation $A=2T_5/t$
between the amplitude $A$ and the amplitude $T_5$ of
Refs.~\cite{babu98,lvov97} taken at the backward angle. The amplitude
$T_5$ contributes to the full Compton scattering amplitude as $T_{fi}
\sim \bar u'\gamma_5 u \,T_5$, and the coefficient of $T_5$ changes
sign when the $P$-operator is applied to the nucleons. The constraint
for $J$ follows from the Landau--Yang theorem \cite{landau} which
states, in particular, that any two-photon system of negative parity
has an even $J$.

Among the lightest hadronic states which have the correct quantum
numbers (\ref{jpc}) to contribute to ${\rm Im}_t A$ are the
pseudoscalar mesons $\pi^0$, $\eta$, and $\eta'$.  Any two-body systems
of pseudoscalar mesons like $\pi\pi$, $K\bar K$, $\pi\eta$, or
$\pi\eta'$ are strictly excluded by parity and the evenness of $J$.
The $3\pi$ continuum is allowed.  Due to $C=+1$ and $I\le 2$, it
necessarily carries isospin $I^G=1^-$ and is produced in the reaction
$2\gamma\to 3\pi$ through anomalous Wess-Zumino-Witten vertices.  The
$3\pi$ continuum appears partly in the form of quasi-two-particle
states like $\pi^0 \sigma$ with even orbital momentum $l$ or
$(\pi\rho)_{I=1}$ with odd $l$ and even $J$, and partly in the form of
broad resonances like $\pi(1300)$.  Furthermore, there is a $4\pi$
continuum, and so on.

\section{Saturation of the sum rule}

The dominating contribution to $\gamma_\pi^s$ comes from single-pion
photoproduction $\gamma N\to\pi N$, which yields the cross section
\beq
   \Delta\sigma^{\pi N} = 8\pi\frac{q^*}{\omega^*}
   \sum_{l=0}^\infty  (-)^l (l+1)
  \Big\{ |A_{l+}|^2 - |A_{(l+1)-}|^2 - \frac{l(l+2)}{4}
   \Big( |B_{l+}|^2 - |B_{(l+1)-}|^2 \Big) \Big\}.
\eeq
Here a sum over channels with charged and neutral pions is implied,
$q^*$ and $\omega^*$ are the pion and photon momenta in the CM frame,
and $A_{l\pm}$ and $B_{l\pm}$ are the standard Walker photoproduction
multipoles \cite{walk69}.  The latter are taken from the computer code
SAID, solution SP97K \cite{arnd96}, and for large angular momenta
$j=l\pm\frac12 \ge \frac92$ from the one-pion-exchange diagram
(Ref.~\cite{lvov97}, appendix B).%
\footnote{Formula (B12) for $f_4$ in that appendix contains a mistake
which is not present in the computer code {\scriptsize\sf GNGN} used to
get numerical results in Ref.~\cite{lvov97}. Instead of the pion
velocity $v$, the ratio of the CM momenta, $q^*/\omega^*$, should stand
there.}
~In Fig.~1 we have plotted the integrand of Eq.~(\ref{SR}) found with
the SAID multipoles.  The integrand is clearly peaked at low energies
and practically vanishes above $\sim 0.5\GeV$, thus supporting a
convergence of the backward dispersion relation.  In order to show
possible uncertainties in evaluating $\Delta\sigma$, we also present in
Fig.~1 results found with the $E_{0+}$, $M_{1-}$, $E_{1+}$, $M_{1+}$
multipoles of Ref.~\cite{hans98} (HDT), which were obtained using
fixed-$t$ dispersion relations and several coupling parameters adjusted
to fit recent experimental data from Mainz and Bonn.  In the HDT set of
multipoles, $A_{0+}$ ($=E_{0+}$) for charged pions near the pion
threshold is larger than that in the SAID set and is closer to
predictions of low-energy theorems. The difference between results
found with the SAID and HDT sets yields an estimate of experimental
uncertainties in the $s$-channel integral.

The evaluation of other (mainly multi-pion) contributions to
$\gamma_\pi^s$ has been done in the framework of the simple model of
Ref.~\cite{lvov97}.  This model takes into account inelastic decays of
$\pi N$ resonances to the channels $2\pi N$, $3\pi N$, $\eta N$, etc.,
and an incoherent nonresonant background consisting of a $\gamma
N\to\pi\Delta$ contribution (calculated in the Born approximation) and
an additional $E1(j=3/2)$ contribution which was adjusted to reproduce
the total photoabsorption cross section. The $s$-wave photoproduction
of $\pi\Delta$ yields a visible negative contribution to $\Delta\sigma$
near the two-pion threshold. The whole integrand related with the
multi-pion states is depicted in Fig.~1.  It is relatively small and
evidently appears to converge.

Even though the multipoles $A_{l\pm}$ and $B_{l\pm}$ are known up to
energies $\omega\sim 2\GeV$, the computation of $\Delta\sigma$ due to
multi-pion channels becomes very model-dependent when $\omega \agt
1\GeV$.  For this reason we cut integrations of all the partial cross
sections in Eq.~(\ref{SR}) at $\omega_{\rm max} = 1.5\GeV$.  The
corresponding integrals are given in Table~1.  They change very little
when a lower cutoff, such as $\omega_{\rm max} = 1\GeV$, is chosen.
Uncertainties in our predictions for $\gamma_\pi^s$ come mainly from
the $\pi N$ channel which results in errors $\simeq \pm 1$, whereas
errors in the multi-pion contribution are estimated to be $\simeq \pm
0.1$.

Next we consider an evaluation of Eq.~(\ref{t}).  Since the $2\pi$
contribution to ${\rm Im}_t A$ is forbidden and since the phase space
for $3\pi$ at low energies is small, one can expect a strong dominance
of $\gamma_\pi^t$ by the lightest pseudoscalar meson $\pi^0$ and, to a
lesser extent, by $\eta$ and $\eta'$.  In the mass region $\sqrt{t}\alt
1.5\GeV$, these mesons and their ``radial excitations" $\pi(1300)$,
$\eta(1295)$, $\eta(1440)$ with $J^{PC}=0^{-+}$ are the only
$t$-channel resonances having the allowed quantum numbers (\ref{jpc}).

Each of the pseudoscalar mesons $M$ makes a contribution
\beq
\label{AM}
  A^{(M)}(t)= \frac{g_{MNN} F_{M\gamma\gamma}}{t-m_M^2}\tau_M
\eeq
to the amplitude (\ref{A}) and, due to the identity
\beq
\label{AM-DR}
  A^{(M)}(t) = \frac1\pi \int_{t_0}^\infty
     {\rm Im}_t A^{(M)}(t')\frac{dt'}{t'-t-i\epsilon},
\eeq
each makes a contribution
\beq
\label{M}
   \gamma_\pi^{(M)} = -\frac{A^{(M)}(0)}{2\pi m}
   = \frac{g_{MNN} F_{M\gamma\gamma}} {2\pi m_M^2 m} \tau_M
\eeq
to the spin polarizability.  Here, the couplings $g_{MNN}$ and
$F_{M\gamma\gamma}$ are those which stand in the effective Lagrangian
\beq
  {\cal L}_{\rm eff} = ig_{MNN} \bar\psi \gamma_5 \tau_M M \psi +
        \frac18 F_{M\gamma\gamma} \epsilon^{\mu\nu\alpha\beta}
        F_{\mu\nu} F_{\alpha\beta} M ,
\eeq
where $\gamma_5= +(i/24)\epsilon^{\mu\nu\alpha\beta} \gamma_\mu
\gamma_\nu \gamma_\alpha \gamma_\beta$, $F_{\mu\nu}$ is the
electromagnetic field tensor, and the isospin factor $\tau_M$ is either
1 or $\tau_3$ for isoscalar and isovector mesons, respectively.  The
radiative couplings $F_{M\gamma\gamma}$ can be fixed through the
two-photon radiative widths $\Gamma_{M\to 2\gamma}=m_M^3
F_{M\gamma\gamma}^2/64\pi$, which are known experimentally.
Alternatively, the ratios
\begin{mathletters}
\label{gammas}
\beqn
  \frac{F_{\eta\gamma\gamma}}{F_{\pi\gamma\gamma}} &=&
    \frac{\cos\theta_P}{\sqrt 3} -
      \sqrt{\frac83}\sin\theta_P \simeq 0.85
      \quad\mbox{(exp.: $0.95\pm 0.06$)},
\\
  \frac{F_{\eta'\gamma\gamma}}{F_{\pi\gamma\gamma}} &=&
    \frac{\sin\theta_P}{\sqrt 3} +
      \sqrt{\frac83}\cos\theta_P \simeq 1.51
      \quad\mbox{(exp.: $1.24\pm 0.07$)}
\eeqn
\end{mathletters}
can be taken from the constituent quark model (CQM). Here $\theta_P
\simeq -10.1^\circ$ is the SU(3) octet-singlet mixing angle for
pseudoscalar mesons.  The strong couplings are reliably known only for
pions, and one can use the value $g_{\pi NN}^2 /4\pi = 13.75$ suggested
by the VPI group \cite{arnd94} which lies between the two extremes
advocated by the Nijmegen \cite{stok93} and Uppsala \cite{eric95}
groups.  The knowledge of strong couplings for $\eta$ and $\eta'$ is
much more uncertain, and we use here predictions of the CQM,
\begin{mathletters}
\label{etas}
\beqn
\label{etaNN}
  \frac{g_{\eta NN}}{g_{\pi NN}} &=&
    \frac{\sqrt 3}{5}\cos\theta_P -
    \frac{\sqrt 6}{5}\sin\theta_P \simeq 0.43,
\\
  \frac{g_{\eta' NN}}{g_{\pi NN}} &=&
    \frac{\sqrt 3}{5}\sin\theta_P +
    \frac{\sqrt 6}{5}\cos\theta_P \simeq 0.42.
\eeqn
\end{mathletters}
The CQM value (\ref{etaNN}) is $\simeq 2.5$ times higher than an
estimate \cite{tiat94} obtained from data \cite{krus95} on
near-threshold $\eta$-photoproduction from the proton.  This higher
$\eta NN$ coupling was found \cite{tiat94} to give, through the Born
diagrams of $\gamma N\to\eta N$, too large $p$-wave photoproduction
multipoles compared to those inferred from the data \cite{krus95}.
Note, however, that the intermediate nucleons in those Born diagrams
propagate very far from the mass shell, and therefore the $\eta NN$
coupling which is effective there is certainly reduced by a nucleon
off-shell form factor in comparison with its on-shell magnitude.  That
is why we suppose that the CQM prediction is not incompatible with the
$\eta$-photoproduction data and use the theoretical estimates
(\ref{etas}) for on-shell nucleons. A cautious reader can take our
estimate as an upper limit.

We note that the CQM predictions for individual strong and
electromagnetic couplings of $\eta$ and $\eta'$ depend on the mixing
angle $\theta_P$ which is not perfectly known. In our estimates we use
a standard choice $\theta_P=-10.1^\circ$ derived from the nonet mass
splitting, athough experimental data on two-photon decays of these
mesons are more compatible with a larger $\theta_P\simeq -20^\circ$
\cite{pdg96}.  However, a combined $t$-channel-exchange contribution of
$\eta$ and $\eta'$ to Compton scattering is not so sensitive to
$\theta_P$.  It would not depend on $\theta_P$ at all in the limit of
equal masses of $\eta$'s.  Including form factors (see below),
$\gamma_\pi^{(\eta)} + \gamma_\pi^{(\eta')}$ varies between $-1.6$ and
$-1.9$ at $\theta_P=-10.1^\circ$ and $\theta_P=-20^\circ$,
respectively.

Taking into account the radial excitations like $\pi(1300)$,
$\eta(1295)$, $\eta(1440)$, etc.\ involves a few more coupling
constants.  We avoid an explicit consideration of these couplings and
assume instead that the radial excitations only renormalize the
contributions (\ref{AM}) of the low-lying mesons $M=\pi$, $\eta$,
$\eta'$ by $t$-dependent form factors which effectively give a $t$
dependence to the couplings in Eq.~(\ref{AM}).  Such a $t$ dependence
does not violate the validity of the dispersion relation (\ref{AM-DR})
for the modified amplitudes $A^{(M)}(t)$, provided the form factors
have singularities (poles or cuts) only at finite real $t>t_0$. These
singularities also contribute to ${\rm Im}_t A^{(M)}$ and, in
accordance with Eq.~(\ref{AM-DR}), result in replacing the original
couplings in Eq.~(\ref{M}) by their magnitudes at $t=0$, $g_{MNN}\to
g_{MNN}(0)$ and $F_{M\gamma\gamma}\to F_{M\gamma\gamma}(0)$.  Thus, we
finally write
\beq
  \gamma_\pi^{t} \simeq \sum_{M=\pi,\eta,\eta'}
   \frac{g_{MNN}(0) F_{M\gamma\gamma}(0)} {2\pi m_M^2 m} \tau_M.
\eeq
Using the experimental values of $g_{\pi NN}$ and $F_{\pi\gamma\gamma}$
known at $t=m_\pi^2$ (cf.\ Eq.~(\ref{pi0}) and Ref.~\cite{lvov97}),
reducing each of them by a monopole form factor $F(0)/F(m_\pi^2)$ with
the cutoff $\Lambda\simeq 1\GeV$, and taking the CQM ratios
(\ref{gammas}), (\ref{etas}) to evaluate the couplings of $\eta$'s at
$t=0$, we obtain the numbers given in Table~1.  As expected, the
largest contribution comes from $\pi^0$ exchange, whereas $\eta$ and
$\eta'$ introduce only a small correction.  With our assumptions for
the form factors, the radial excitations play a minor role, reducing
for example the contribution from $\pi^0$ by only $1 - F^2(0) /
F^2(m_\pi^2) \simeq 4\%$, thus again supporting a convergence of the
sum rule.  A conservative estimate of uncertainties in $\gamma_\pi^t$
is $\simeq \pm 1.6$ due to $\pi^0$, another $\simeq \pm 1$ due to
$\eta$'s, and an additional $\simeq \pm 1$ due to the form factors.

In total, gathering all the $s$- and $t$-channel pieces and combining
uncertainties in quadrature, we obtain the non-$\pi^0$ part of the
backward spin polarizability of the proton $[\gamma_\pi^{(\rm
non-\pi^0)}]_p \simeq 5.5 \pm 1.8$.  This number is only $\simeq
\frac13$ of the experimental estimate (\ref{exp0}).  Both for the
proton and the neutron, $\gamma_\pi^{(\rm non-\pi^0)}$ is dominated by
the $s$-channel contribution (\ref{s}) from low energies $\omega \alt
500$ MeV, which is well under control, whereas the higher energies and
the heavier $t$-channel exchanges contribute little (see Fig.~1 and
Table~1 for more details).  Therefore, it is difficult to reconcile our
result with the experimental finding.

\section{Conclusions}

Using the backward-angle dispersion relation, we derived a novel
well-convergent sum rule for the backward spin polarizability.  We
applied it for an evaluation of $\gamma_\pi$ for the nucleon and
obtained results which were numerically close to both the naive
dispersion estimates \cite{drec98,babu98} based on an unsubtracted
fixed-$t$ dispersion relation for $A_2$ and the leading-order results
of ChPT with the $\Delta(1232)$ included \cite{hols98}.  Our results
support the physical conclusion inferred from the previous work that
$\gamma_\pi$ is strongly dominated by low-energy subprocesses of pion
photoproduction, including the $\Delta$-isobar excitation, and by
$\pi^0$ exchange.  They disagree, however, with the experimental
finding of Ref.~\cite{legs98} which indicates a large additional
contribution, perhaps from higher energies.  Therefore, we suggest that
the indirect estimate (\ref{exp}) be checked in further dedicated
experiments with polarized particles.  Some possibilities for that are
sketched in Ref.~\cite{babu98}.

\acknowledgements

A.L. acknowledges the hospitality of the Institut f\"ur Kernphysik at
Mainz and of the University of Illinois at Urbana-Champaign, where a
part of this work was done.  The work was supported in part by the
Russian Foundation for Basic Research, grant 98-02-16534, and by the
DFG (SFB 201).

\begin{figure}[htb]

\epsfig{file=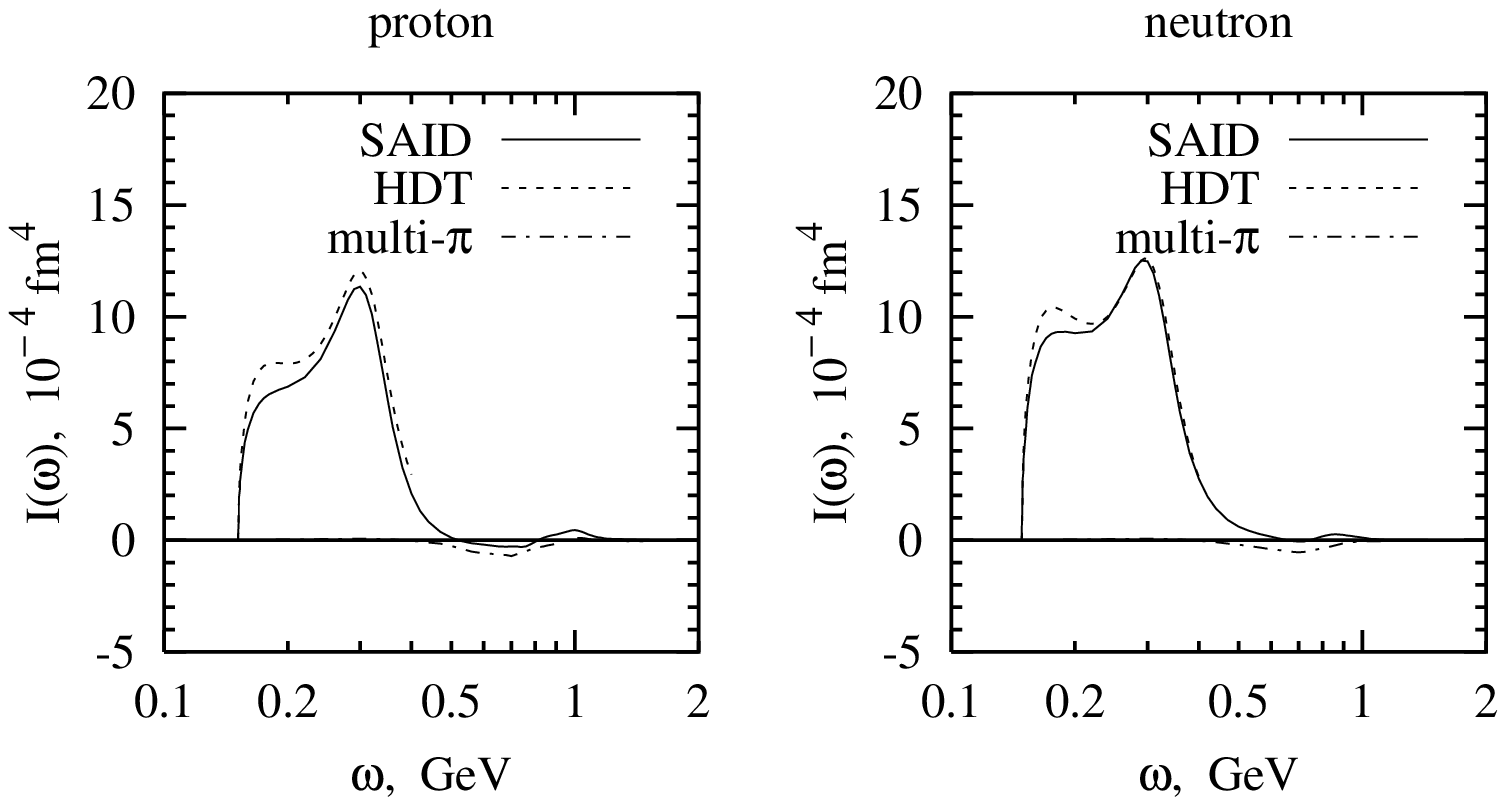}

\caption{The $s$-channel integrand of
 $\gamma_\pi^s = \displaystyle\int_{\omega_0}^\infty
     I(\omega)\,d\omega/\omega$,
 ~$I(\omega)=
  \displaystyle \protect\rule[-3ex]{0ex}{6ex}
 \frac{1}{4\pi^2\omega^2}\sqrt{1+\frac{2\omega}{m}}
   \Big(1+\frac{\omega}{m}\Big)  \Delta\sigma$
for the proton and neutron.
Solid lines: the contribution of $\gamma N\to\pi N$ with the SAID
multipoles \protect\cite{arnd96}.  Dashed lines: the same with the HDT
multipoles \protect\cite{hans98}.  Dashed-dotted lines:  the
contribution of multi-pion states.}

\end{figure}

\begin{table}[htb]
\vspace{3em}
\begin{tabular}{lrr}
\qquad $10^{-4} ~\rm fm^4$   &    proton    &    neutron   \\
\hline
  $s$-channel, $\gamma N\to\pi N$                       && \\
\qquad SAID, 150--500 MeV    &  $  7.29  $  &  $  9.22  $  \\
\quad\qquad  (HDT, 150--500 MeV)
                             &  $ (8.35)\!\! $  &  $ (9.63)\!\! $  \\
\qquad SAID, ~500--1500 MeV  &  $  0.02  $  &  $  0.13  $  \\
  $s$-channel, $\gamma N\to(\ge 2\pi) N$
                             &  $ -0.28  $  &  $ -0.23  $  \\
\hline
  $t$-channel                                           && \\
\quad  $\pi^0$               &  $ -45.0 \pm 1.6  $  &  $ +45.0  \pm 1.6 $  \\
\quad  $\eta$                &  $ -1.00  $  &  $ -1.00  $  \\
\quad  $\eta'$               &  $ -0.57  $  &  $ -0.57  $  \\
\hline
  total non-$\pi^0$          &  $   5.5 \pm 1.8  $  &  $   7.5  \pm 1.8  $  \\
  total                      &  $ -39.5 \pm 2.4  $  &  $  52.5  \pm 2.4  $  \\
  experiment \cite{legs98}   &  $ -27.1 \pm 3.4 $
\end{tabular}

\bigskip
\caption{The backward spin polarizability of the nucleon,
Eq.~(\protect\ref{SR}). Separately shown are $s$-channel integrals from
single-pion photoproduction with SAID \protect\cite{arnd96} and HDT
\protect\cite{hans98} multipoles, and from multi-pion production.  The
$t$-channel integral includes both the pole contributions of low-lying
pseudoscalar mesons and their radial excitations (see the text).}
\end{table}

\end{document}